\documentclass[12pt]{article}

\usepackage{amssymb}

\usepackage{graphicx}
\usepackage{authblk}
\usepackage{physics}
\usepackage{xcolor}

\newcommand{\vc}[1]{\boldsymbol{#1}}
\newcommand{\vt}[1]{\boldsymbol{\mathsf{#1}}}
\newcommand{\vv}[1]{\mathsf{\boldsymbol{#1}}}

\DeclareFontEncoding{LS1}{}{}
\DeclareFontSubstitution{LS1}{stix}{m}{n}
\DeclareSymbolFont{symbols4}{LS1}{stixbb}{m}{it}
\DeclareMathSymbol{\varhexagonblack}{\mathord}{symbols4}{"DD}
\DeclareMathSymbol{\hexagonblack}   {\mathord}{symbols4}{"DE}

\begin{document}



\title{Smoothed Particle Hydrodynamics simulations of integral multi-mode and fractional viscoelastic models}
\date{}
%

\author[1]{Luca Santelli}
\affil[1]{Basque Center for Applied Mathematics (BCAM),
        Alameda de Mazarredo 14, 
           Bilbao,
            48009, 
           Bizkaia,
            Spain}
\author[2]{Adolfo V\'azquez-Quesada}
\author[1,3,4]{Marco Ellero}
\affil[2]{Departamento de Fisica Fundamental, UNED,
            Apartado 60141, 
            Madrid,
            28080, 
            Spain}

\affil[3]{IKERBASQUE, Basque Foundation for Science,
            Calle de María Díaz de Haro 3, 
            Bilbao,
            48013, 
            Bizkaia,
            Spain}           
           
\affil[4]{Zienkiewicz Center for Computational Engineering (ZCCE), Swansea University,
            Bay Campus, 
            Swansea,
            SA1 8EN, 
            United Kingdom}
\renewcommand\Affilfont{\itshape\small}

\maketitle
\begin{abstract}

To capture specific characteristics of non-Newtonian fluids, during the past years fractional constitutive models have become increasingly popular. These models are able to capture in a simple and compact way the complex behaviour of viscoelastic materials, such as the change in power-law relaxation pattern during the relaxation process of some materials. Using the Lagrangian Smoothed--Particle Hydrodynamics (SPH) method we can easily track particle history; this allows us to solve integral constitutive models in a novel way, without relying on complex tasks.


Hence, we develop here a SPH integral viscoelastic method which is first validated for simple Maxwell or Oldroyd-B models under Small Amplitude Oscillatory Shear flows (SAOS). By exploiting the structure of the integral method, a multi-mode Maxwell model is then implemented. Finally, the method is extended to include fractional constitutive models, validating the approach by comparing results with theory under SAOS.

\end{abstract}


%
%


\section{Introduction}
It is very common to approach the study of flows of viscoelastic fluids using  differential constitutive models, in which the stress evolution only depends on the current velocity and stress fields. Nevertheless, there are complex systems, such as many polymer melts, for which an integral formulation of the constitutive equations provides a more realistic representation of their fluid behaviour \cite{bird1987dynamics, doi1988theory, tanner2000engineering, owens2002computational}. 

Numerically, it would be preferable to study these complex fluids by using a numerical method that incorporates the integral formulation without introducing additional approximations, which is often the case when one tries to implement a constitutive equation via a differential discretisation \cite{larson1999structure}. Researchers have been working on the topic for nearly fifty years, starting with the works of \cite{viriyayuthakorn1980finite} and developing several different schemes, mostly focusing on finite element \cite{dupont1985finite, henriksen1994parallel,hulsen2001new,keunings2003finite}  and finite difference methods \cite{tome2016finite}.
However, integral models provide a series of challenges which are not easy to deal with. Mainly, any numerical implementation must deal with three tasks: (i) solve nonlinear iterative process, (ii) deal with the constitutive equation written in a Lagrangian form, and (iii) evaluate the convolution of the strain with a memory function along particle paths which are, a priori, unknown. The latter is often described as the most complex task, since in order to compute the required Cauchy-Green and Finger strain tensors, it is necessary to have a way to track particle history among the whole fluid domain \cite{aggarwal1994simulation, luo1988finite}. This usually requires performing computationally expensive procedures to compute the past positions of particles, and -to the authors knowledge- is the main reason integral models have not seen an increase in popularity over the last years.

Smoothed Particle Hydrodynamics (SPH) is a Lagrangian mesh-free numerical method that solves evolution equations for fluids by introducing interacting fluid particles distributed over the domain, and interpolating flow properties over these particles. This method, originally developed for astrophysics \cite{lucy1977numerical, gingold1977smoothed}, has been applied with success to describe many different physical systems \cite{morris1997modeling, ellero2007incompressible, ellero2010sph}. Studies, in the last two decades, have shown that SPH is very effective in simulating viscoelastic non-Newtonian fluids  \cite{ellero2002viscoelastic,ellero2005sph, xu2016multiscale, king2021high}. This finding has opened the path to several studies in the matter of complex fluids, including thermal fluctuations in microrheology \cite{vazquezquesada2009smoothed, vazquezquesada2012sph}, suspension of particles in Newtonian and non-Newtonian solutions \cite{vazquezquesada2016three, vazquezquesada2017sph, vazquezquesada2019shear}, and magnetic suspensions \cite{rossi2021dynamics,lagger2015influence}. SPH has found applications also in the field of viscoplasticity, with studies on debris and granular flows \cite{rodriguez2004corrected,zhu2010numerical,minatti2015sph}, multiphase flows \cite{hu2006angular, fourtakas2016modelling}, and interaction fluid-structure \cite{zhu2018sph}, as well as thixotropic fluids \cite{rossi2022sph}. It has been successfully applied to the study free surface flows on Maxwell, Oldroyd-B and Cross model fluids with a focus on tensile instability, impacting drop, jet buckling and 3D injection molding \cite{fang2006numerical,fang2009improved, rafiee2007incompressible, xu2013sph, xu2022sph}, and also used in simulation of 3D dispersion of sediments \cite{tranduc2019three} and fiber-filled composites in non-isothermal three-dimensional printing processes \cite{ouyang2019smoothed}. 
The flexibility of the method allowed for a dissipative implementation of SPH that has been used to perform multiscale simulations of Newtonian and non-Newtonian fluids \cite{moreno2023generalized}; moreover, 
 simple and complex fluids (e.g. Oldroyd-B fluids and polymeric solutions) have been simulated via SPH
 in periodic array of cylinders \cite{vazquezquesada2012spha,grilli2013transition, simavilla2022mesoscopic} and extrudate swell \cite{xu2016improved}.



All these implementations relied on differential constitutive equations to describe the physics of the fluid. Nevertheless, SPH is particularly effective in dealing with integral constitutive equations; indeed, thanks to its Lagrangian nature and formulation, particle history is an immediately available information which does not require additional steps to compute.  Our goal is to exploit this structure of SPH to describe the general behaviour of a range of viscoelastic fluids. In the integral formulation, the physical modelling is completely included in the memory kernel of the integral; therefore, once a numerical method has been developed for a particular choice of a memory function, it is extremely easy to extend the method to different constitutive equations.

The specific physical case we are interested in exploring is a polymer melt under small amplitude oscillatory shear (SAOS), i.e. the behaviour of the fluid when posed between two parallel planes, one of which oscillating with a fixed frequency $\omega$. A good description for this complex fluid is often achieved via a non linear generalization of the Maxwell model, e.g. models based around an exponential behaviour in the stress relaxation. However, this description is non suitable for SAOS, where the small amplitude implies a linearity of the solution, and all the nuances obtained by adding non linear terms to the Maxwell model are lost in this regime. Indeed, polymer melts belong to the class of material for which the stress relaxation after a step strain does not follow an exponential behaviour in time, and is best represented by a power-law decay. In this class of power-law materials fall many complex fluids, such as microgel dispersions \cite{ketz1988rheology}, soft glassy materials \cite{sollich1998rheological}, and colloidal hard sphere suspensions \cite{mason1995linear}.  This power-law behaviour in the stress relaxation translates directly in a similar behaviour for the elastic $G'(\omega)$ and loss $G''(\omega)$ moduli, which are the characterising rheological quantities under SAOS.


It is well known that a power-law curve can be fitted by a linear combination of exponential functions \cite{moore1974exponential,sastry2012introductory}.
In the study of viscoelasticity, this means creating a model that includes multiple Maxwell elements, each with its own mode of decay, to obtain the so called multi-mode (or generalised) linear Maxwell model \cite{bird1987dynamics,tschoegl1989phenomenological,winter1986analysis}. However, although it is, in principle, possible to increase the number of modes to reproduce any complex rheological behaviour on a prescribed frequency window, this ad-hoc formulation might become computationally expensive when a large number of modes is needed, and fitting the model to experimental data will provide parameter values that have a strong dependence on the time scale of the experiment, suggesting a lack of physical meaning in the obtained parameters \cite{kollmannsberger2011linear}. Moreover, the number of modes required scales with the orders of magnitude of frequency explored, making this approach harder to use when a broader range of frequency is studied \cite{tanner2000engineering}. 

An alternative approach, that naturally produces a power-law behaviour, is using fractional constitutive models \cite{gemant1936method,gemant1938fractional, scottblair1947role}. These models introduce a constitutive element known as the spring-pot, that interpolates between the response of a spring and a dashpot (i.e. the constitutive elements of the Maxwell model) \cite{jaishankar2013power,jaishankar2014fractional, yang2010constitutive}. Each spring-pot is characterised by two parameters, an exponent $0\le\alpha\le 1$, controlling the time scale, and a quasi-property $\mathbb V$ controlling the magnitude.
A single spring-pot is suited to describe a class of materials known as critical gels \cite{winter2002critical}, which exhibit a pure power-law decay $G(t)\propto \mathbb V t^{-\alpha}$. More complex behaviours, especially for materials with multiple time scales, can be captured by adding multiple spring-pot in series or in parallel: the Fractional Maxwell Model is constructed by adding two spring-pots in series and can accurately describe polymer melts. The constitutive equation associated with the spring-pot requires the use of the mathematical concept of fractional derivatives, which have a long history of studies in fractional calculus but have been only recently applied to viscoelasticity \cite{rathinaraj2021incorporating, sharma2010polymer, almusallam2023large}. 

In this work we propose a novel SPH-based integral model able to describe both multimode viscoelastic as well as fractional models in a very flexible way.
This manuscript is organised as follow: in section \ref{sec:integral_model} we will discuss the mathematical details of the integral model, starting from its generic formulation and then delving into fractional constitutive models; the numerical implementation of the integral approach into Smoothed Particles Hydrodynamics will be discussed in section \ref{sec:SPH}; and the numerical simulation results will be shown in section \ref{sec:results}. The findings and the future directions of this work are discussed in section \ref{sec:conclusions}.

\section{Integral Viscoelastic Models}
\label{sec:integral_model}

We will introduce now the variables required for building an integral model, following closely the approach of \cite{bird1987dynamics}, which is a strongly suggested reference for any further in-dept analysis.
The main variable is the displacement function $\vb{r} = \vb{r}(\vb r', t', t)$, which tracks the current position $\boldsymbol{r}$ of a particle, at time $t$, as a function of its past position $\vb r'$ at time $t'$. In this representation, the pair $\vb r', t'$ identifies the particle we are tracking. \textcolor{black}{The Cartesian components of $\boldsymbol{r}$} are identified by $x_\mu$, with the Greek index ranging between the number of dimensions.
An alternative and mathematically equivalent function is $\vb r'=\vb r'(\vb r,t,t')$, which tracks the past position at time $t'$ of a particle identified by its current position $\vb r$ at time $t$. In our formalism we will mostly focus on $\vb r$, however its counterpart will appear in some of the definitions.


From the displacement function we define a displacement gradient
\begin{equation}
    \Delta_{\mu\nu} (\vb r , t, t') = \pdv{x'_\mu(\vb r,t,t')}{x_\nu}
    \label{eq:grad_r'_bird}
\end{equation}
i.e. the gradient of the past position of the fluid particles at time $t'$ around the current position $\vb r$.
The equivalent form in terms of displacement of the current position around the previous is 
\begin{equation}
    E_{\mu\nu} (\vb r , t, t') = \pdv{x_\mu(\vb r',t',t)}{x'_\nu}.
\end{equation}
Together, they satisfy $\sum_{\mu,\nu}\Delta_{\mu\nu}E_{\mu\nu}=\delta_{\mu\nu}$ and, for incompressible fluids, $\det(\Delta_{\mu\nu})=\det(E_{\mu\nu})=1$.
However, these are not the best candidates to use for a constitutive model, since they are generally not symmetrical and are not invariant under rigid body motions.
Therefore we need to introduce two finite strain tensors, Cauchy ($\vt B^{-1} $) and Finger ($\vt B$):
\begin{align}
&B^{-1}_{\mu\nu}(\vb r,t,t') = \sum_\xi\pdv{x'_\xi}{x_\mu}\pdv{x'_\xi}{x_\nu}\\
&B_{\mu\nu}(\vb r,t,t') = \sum_\xi\pdv{x_\xi}{x'_\mu}\pdv{x_\xi}{x'_\nu}
\end{align}
They satisfy the same properties shown above of $\vb \Delta$ and $\vb E$; moreover, they are symmetrical, positive definite, and invariant under rigid rotation and translation. Thus they can be used to build finite strain tensors that vanish for rigid body motions:
\begin{align}
    &\vt\gamma^{[0]} = \vt B^{-1}-\vt 1\\
    &\vt\gamma_{[0]} = \vt 1 - \vt B
\end{align}
where $\vt 1$ is the identity tensor and$\boldsymbol{\gamma}^{[0]}$ and $\boldsymbol{\gamma}_{[0]}$ are the so-called relative finite strain tensors. 
We note that, while for small displacements $\vt\gamma^{[0]}=\vt\gamma_{[0]}=\vt\gamma$, with $\vt\gamma$ being the infinitesimal strain tensor, this is, in general, not true for large displacements. Comparison with experimental data shows that more realistic constitutive models, e.g. the corresponding integral version of the upper convected Maxwell,  are built upon the finite strain tensor $\vt\gamma_{[0]}$
\cite{bird1987dynamics, yang2010constitutive, dengke2005exact, lodge1964elastic}. However, since we are interested in the behaviour under SAOS, i.e. linear small deformations, we can exploit the equivalence between $\vt\gamma$ and $\vt\gamma_{[0]}$ in the following. Nevertheless, the proposed methodology will be completely general and can be applied easily to the case of both $\vt\gamma^{[0]}$ and $\vt\gamma_{[0]}$.

Finally, we note that in the building of the integral model there is no dependence on the velocity field. If such a field is needed, it can be directly derived by the displacements.
\begin{align}
    \vb v'(\vb r',t') &= \pdv{\vb r'(\vb r,t,t')}{t'}\\    
    \vb v(\vb r,t) &= \pdv{\vb r(\vb r',t',t)}{t}
\end{align}

To build the constitutive model, we continue following the approach of \cite{bird1987dynamics} and start from the linear Maxwell model which, in differential form, reads 
\begin{equation}
    \vt \tau + \lambda_1 \pdv{\vt \tau}{t} = -\eta_0\dv{\vt\gamma}{t}
\end{equation}
where $\vt \tau$ is the stress, $\lambda_1$ is the characteristic relaxation time, $\eta_0$ is the zero-shear viscosity and $\dot{\vt\gamma}\equiv\dd\vt\gamma/\dd t$ is the strain rate. 
Solving for the stress tensor gives
\begin{equation}
    \vt \tau(t) = -\int_{-\infty}^t\qty{\frac{\eta_0}{\lambda_1}\exp(-\frac{t-t'}{\lambda_1})}\dot{\vt \gamma}(t')\dd t'
\end{equation}
The quantity between curly braces is called the relaxation modulus $G(t-t')$.
Noting that $\vt \gamma(t,t')=\int_t^{t'}\dot{\vt\gamma}(t'')  \dd t''$ for small displacements, by integrating by part we obtain 
\begin{equation}
    \vt \tau(t) = \int_{-\infty}^t\qty{\frac{\eta_0}{\lambda_1^2}\exp(-\frac{t-t'}{\lambda_1})}\vt\gamma(t,t')\dd t
    \label{eq:memory_singleMaxwell}
\end{equation}
 and the quantity between curly braces, called the memory function $M(t-t')$, satisfies 
 \begin{eqnarray}
     M(t-t')=\pdv{G(t-t')}{t'} .
     \label{eq:memory_vs_relaxation}
 \end{eqnarray}
Different constitutive model can be built by choosing different memory/relaxation functions. Therefore 
 \begin{equation}
    \vt \tau(t) = \int_{-\infty}^tM(t-t')\vt\gamma(t,t')\dd t
    \label{eq:memory_integral}
\end{equation}
 can be interpreted as a generic integral constitutive equation. 
 Equation \eqref{eq:memory_integral} can then be referred to a specific fluid model by an appropriate choice of the memory function. 
For example, the multi-mode Maxwell fluid model is obtained with 
 \begin{equation}
 M(t-t') = \sum_k\frac{\eta_k}{\lambda_k^2}\exp(-\frac{t-t'}{\lambda_k})
 \label{eq:memory_multiMaxwell}
 \end{equation}
 with $\eta_k$ and $\lambda_k$ the polymeric viscosity and relaxation time for each mode $k$, and the sum done over all the modes. Thanks to this formulation, numerical studies using different constitutive equations require little effort once the first scheme has been set up, indeed the only necessary requirement is the choice of the memory function, while everything else follows the same structure.
 \subsection{Fractional viscoelasticity}
\begin{figure}
     \centering
     \includegraphics[width=0.8\linewidth]{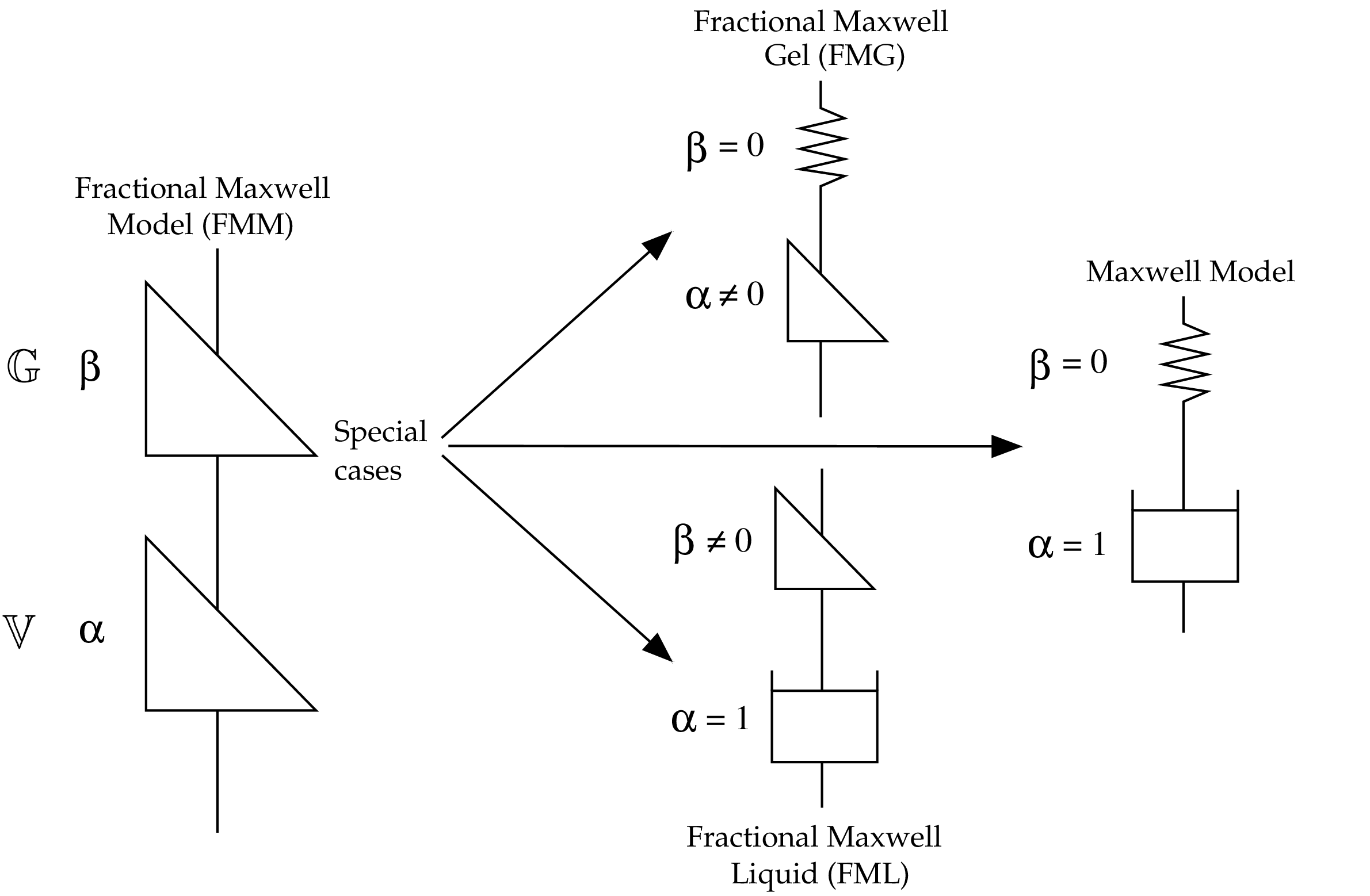}
     \caption{Two spring-pots in series, forming the Fractional Maxwell model. On the left, generic parameters $\alpha, \beta\, \mathbb V, \mathbb G$. On the right, special cases for $\alpha\to1$ and $\beta\to0$}.
     \label{fig:fractional_maxwell}
 \end{figure}

 The Fractional Maxwell Model consists of two spring-pots, each characterized by a pair of material parameters (here, denoted by $\mathbb V$, $\alpha$ and $\mathbb G$, $\beta$, respectively), arranged in series as shown in figure \ref{fig:fractional_maxwell}.
 
 It can be shown that the spring-pot is a compact representation of a recursive ladder model constructed from Maxwell elements \cite{sadman2017influence,schiessel1995mesoscopic}. The shear stress response of a single spring-pot, characterized by a quasi-property $\mathbb G $ (with units of $Pa\cdot s^\beta$) and an exponent $\beta$, can be written in terms of fractional derivative as 
 \begin{equation}
     \vt\tau(t) = \mathbb{G}\dv[\beta]{\vt\gamma(t)}{t}, 
 \end{equation}
 where $\vt \gamma(t<0)=0$ and we choose to use the Caputo fractional derivative defined as
 \begin{equation}
    \dv[\beta]{\vt\gamma(t)}{t} \equiv \frac{1}{\Gamma(1-\beta)}\int_0^t (t-t')^{-\beta}\dot{\vt\gamma}(t')\dd t' 
 \end{equation}
 with  $\Gamma(\cdot)$ the complete Gamma function and $0\le\beta\le 1$. 
 
Following the procedure of the previous section, by solving for the stress we obtain
\begin{equation}
    \vt\tau(t) = \frac{\mathbb{G}}{\Gamma(1-\beta)}\int_0^t(t-t')^{-\beta}\dot{\vt\gamma}(t')\dd t' = -\frac{\mathbb{G}\beta}{\Gamma(1-\beta)}\int_0^t(t-t')^{-\beta-1}\vt\gamma(t,t')\dd t'
\end{equation}
As we are interested in steady shear flow problems, henceforth we will focus on the $xy$ component of tensors by introducing $\tau\equiv\tau_{xy}$ and $\gamma\equiv\gamma_{xy}$. Applying a step strain $\gamma(t) = \gamma_0 H(t)$ results in a power-law decay of the stress: 
\begin{equation}
    \tau(t)/\gamma_0 \equiv G(t) = \frac{\mathbb{G}}{\Gamma(1-\beta)}t^{-\beta}
    \label{eq:relax_modulus_singleSP_time}
\end{equation}
which is the response of a class of material known as critical gel \cite{rathinaraj2021incorporating}. The behaviour under SAOS can be computed by Fourier transformation of Equation \eqref{eq:relax_modulus_singleSP_time}, obtaining
\begin{align}
    &G'(\omega) = \mathbb{G}\cos(\pi\beta/2)\omega^\beta\\
    &G''(\omega) = \mathbb{G}\sin(\pi\beta/2)\omega^\beta
\end{align}

To incorporate the physics of more complex materials, such as polymers, spring-pots can be arranged in series or in parallel. As anticipated, two spring-pots in series result in the Fractional Maxwell Model, which satisfies
\begin{equation}
    \tau(t) + \frac{\mathbb V}{\mathbb G}\dv[\alpha-\beta]{\tau(t)}{t}=\mathbb V \dv[\alpha]{\gamma(t)}{t}
\end{equation}
As usual, solving for the stress allows us to write the solution in the integral form. Here we report the relaxation modulus and the memory function: 
\begin{align}
    &G(t-t') = \mathbb G (t-t')^{-\beta}E_{\alpha-\beta, 1-\beta}\qty(-\frac{\mathbb G}{\mathbb V}(t-t')^{\alpha-\beta})\\
    &M(t-t') = -\mathbb G (t-t')^{-1-\beta}E_{\alpha-\beta, -\beta}\qty(-\frac{\mathbb G}{\mathbb V}(t-t')^{\alpha-\beta})
    \label{eq:memory_fractional}
\end{align}
where \begin{equation}
    E_{a,b}(z) = \sum_{k=0}^\infty \frac{z^k}{\Gamma(ak+b)}
\end{equation}
is the generalized Mittag-Leffler function. Heuristic arguments in favour of this description for polymeric materials can be found looking at experimental data \cite{jaishankar2014fractional}, while a microscopic explanation is presented in \cite{sharma2010polymer}. 

The Fractional Maxwell model presents some interesting limits, as shown in \ref{fig:fractional_maxwell}. The linear Maxwell model can be seen as a limit case of the fractional Maxwell for $\alpha=1, \beta=0$, i.e. when one spring-pot becomes a dashpot and the other a spring. Other two important limiting cases are the Fractional Maxwell Liquid ($\alpha=1$), which shows a bounded steady shear viscosity $\eta_{FML}=\mathbb V$ and is a good description for fluids in the pre-gel state, and the Fractional Maxwell Gel ($\beta=0$), which captures the elastic behaviour of viscoelastic materials beyond the gel point and for which $G'_{FMG}(\omega\to\infty)=\mathbb G$ \cite{rathinaraj2021incorporating}. 

 \section{SPH implementation}
 \label{sec:SPH}
 
The evolution of the system is discretized via Smoothed Particle Hydrodynamics (SPH), accurately described in \cite{ellero2010sph,  vazquezquesada2009smoothed}. Here we will provide a short summary of the method for the sake of completeness. SPH  is a mesh-free Lagrangian model that discretizes the prescribed equations using a set of fluid particles. Each particle represents a discrete fluid element and is associated to its position and momentum (labeled by $i,j=1,\dots,N_p$). 
The conservation of mass is automatically satisfied by defining a particle number density 
\begin{equation}
d_i\equiv\sum_j W(|\vb r_{ij}|, r_{cut})
\end{equation}
where (here and thereafter) the sum is intended over all the neighbouring particles of particle $i$, $\vb r_{ij} \equiv \vb r_i - \vb r_j$ is the distance between position $\vb r_i$ and $\vb r_j$ and its normalised value is defined as $\vb e_{ij}=\vb r_{ij}/ r_{ij}$. $W(|\vb r_{ij}|, r_{cut})$ is an interpolating kernel. Here we used a quintic spline kernel function with compact support $r_{cut}$, defined in \cite{morris1997modeling}. For this work, a cutoff radius $r_{cut}=4\Delta x$ has been chosen, with $\Delta x$ being the mean SPH particle separation. 
Given the particle mass  $m$, the mass density is $\rho_i=md_i$.
The equation of motion for each particle $i$ is given by a discretization of the Navier-Stokes equations \cite{espanol2003smoothed}, reading
\begin{align}
    \vb{\dot{r}}_i&= \vb v_i\\
%
 %
%
      m\dot{\vb v_i} &= -\sum_j  \qty[\frac{\vc \pi_i}{d_i^2}+ \frac{\vv{\pi}_j }{d_j^2}]W'_{ij}\vb e_{ij}+2\eta_s\sum_j\frac{1}{d_id_j}\frac{W'_{ij}}{r_{ij}} \vb v_{ij}
    \label{eq:SPH_vel}
\end{align}
%
where $\eta_s$ is the viscosity of the Newtonian solvent 
and $W'_{ij}\equiv\pdv{W(\vb r,r_{cut})}{\vb r} \eval_{\vb r = \vb r_{ij}}$. The total stress tensor $\vc \pi_i=P_i\vt{1} -\vt \tau_i$ can be separated in the contributions of the isotropic particle pressure $P_i$ and the extra-stress $\vt \tau_i$. 
In the case of a simple Newtonian fluid, the total stress is therefore simply the fluid pressure multiplied by the identity tensor $\vt 1$, which can be modelled by introducing an equation of state for the pressure $P_i = c_s^2(\rho_i-\rho_0)$, where $\rho_0$ is a reference fluid density and $c_s$ is the speed of sound in the solvent \cite{monaghan1994simulating}. $c_s$ can be chosen based on a scale analysis to keep density variation small enough (weakly compressible approximation). Here, equation \eqref{eq:SPH_vel} can have an immediate connection with Navier-Stokes equations, being the first term on the right-hand side linked to the pressure gradient, while the second term is the discretization of the Laplacian of the velocity, and is related to the viscous dissipation \cite{ellero2010sph}. Polymeric physics can be integrated by a proper choice of the extra-stress. Traditional approaches introduce an additional differential equation for $\vt \tau$ (or other microstructural tensorial variables), based on specific objective derivatives (upper convected, lower convected, Jaumann, etc.) \cite{vazquezquesada2009smoothed,vazquezquesada2019shear,simavilla2023non}. 
However, additional care is required when computing $\vt \tau_i$ following the integral scheme, as we shown in the next section. 

 \subsection{Integral computation of the extra-stress}
 \label{section_integral_compute}
 
In order to numerically compute $\vt \tau$ from equation \eqref{eq:memory_integral}, we must store 
 the history of each computational particle.
 The position $\vb r_i(t)$ of each particle $i$ is readily accessible in SPH at each time step, allowing for direct storage of their history. To address periodic boundary conditions effectively, our program stores particle displacements $\Delta\boldsymbol{r}_i(t,t') = \boldsymbol{r}_i(t) - \boldsymbol{r}_i(t')$, simplifying updates.
 For clarity, we will denote $\Delta \boldsymbol{r}_{i,n} = \boldsymbol{r}_i(t) - \boldsymbol{r}'_i(t-n \Delta t)$, 
 with $\Delta t$ the time step, to refer to the displacement of the particle $i$, $n$ time steps prior to the current time.
 The update of the displacements at time step $(n+1)\Delta t$ reads
 \begin{eqnarray}
  \Delta \boldsymbol{r}_{i,{n+1}} =
  \Delta \boldsymbol{r}_{i,n} + 
  \boldsymbol{r}_i(t+\Delta t) - 
  \boldsymbol{r}_i(t)
  \label{update_displacement}
 \end{eqnarray}
  
To compute the gradient of the position vector field $\boldsymbol{r}'$ at a previous time $t'$, evaluated at the current positions of all particles, we use the following SPH expression for the gradient
  \begin{eqnarray}
    \vt \Delta_i \equiv \qty(\grad \vb r ')_i = -\frac{1}{d_i}\sum_jW'_{ij}\vb e_{ij}\vb r'_{ij} 
\end{eqnarray}
where $\boldsymbol{r}'_i = \boldsymbol{r}_i(t')$ denotes the position of the particle $i$ at the time $t'$ and $\boldsymbol{r}'_{ij} = \boldsymbol{r}'_i - \boldsymbol{r}'_j$. Note that, if $t'$ is the current time, this expression is a SPH version of the identity tensor.
The discretised tensor $\vt E_i$ is then computed as the inverse of $\vt \Delta_i$, and therefore the relative finite strain tensor on particle $i$ is obtained as
\begin{equation}
    \vt \gamma_{[0],i} = \vt 1 - \vt E_i \vt E_i^T.
\label{relative_finite_stress_tensor}
\end{equation}

Thus, once a proper memory function is chosen, the final discretised stress is computed as
\begin{equation}
    \vt \tau_i = \int_{t_0-T_{save}}^{t_0} M(t-t')\vt \gamma_{[0],i}(t,t') \dd t' .
    \label{stress_calculation}
\end{equation}
 
Nevertheless, it is possible to greatly improve the efficiency of the algorithm by
 not considering all the history of the displacements. 
 First, given that the memory function has a monotonically decreasing behaviour, we can store the positions up to a sufficiently large past time $T_{save}\equiv N_s\Delta t$.     
 The typical value in our simulations is $T_{save} = 10\lambda $, where $\lambda$ is the largest elastic relaxation time. Even with this choice, typically, we need to store the positions of the particles for approximately $N_s \sim 10^5$ time steps in our simulations.
While it is possible to store all these positions, it is unnecessary. Furthermore, calculating the gradients of all the previous position fields is computationally expensive, considering that we only need a subset of them for an efficient computation of integral \eqref{eq:memory_integral}. Therefore, we can significantly reduce the amount of saved data by saving only one configuration every $f$ steps, leading to a much smaller number $N_s$ of saved configurations. The accuracy of this approximation depends on the value of the saving frequency $f$, which therefore must be tuned according to the analysed problem. In particular, we want to ensure that the stress changes are captured at the fastest 
time scales, which in the simulations we carried out are represented by the frequency of the oscillation and the elastic relaxation time. In our set-up, this criterion for $f$ leads to $N_s=\order{10^2}$. 
However the time interval $\Delta t_s$ between the last saved configuration and the current time $t_{0}$, is not fixed, and increases until we reach the new threshold for a new update of the positions. Therefore  the corresponding time for each element of the array is $t_n = t_{0}- \Delta t_s - (N_s-l)f\Delta t$, where $l$ is an index labeling each of the saved configurations from $1$ to $N_s$. For additional information on how we handle the save configurations, please refer to \ref{appendix_integration}. This method significantly reduces the amount of data we need to store; however it leaves out the contribution of the integral between the last save configuration and the current time, i.e. at the most recent time interval $\Delta t_s$. This contribution can be approximated analytically. Let's focus on the shape of the Fractional memory function of equation \eqref{fig:fractional_maxwell}. This equation has a weakly divergent contribution for $t'\to t_{0}$, which makes numerical computation harder. For small enough time intervals, we can approximate the velocity gradient at the position of the particle with a constant. Under this approximation, the evolution of the Finger tensor \cite{huilgol1997fluid}
\begin{equation}
       \dot{\vt{B}} = \nabla\vb{v}\cdot \vt{B}
\end{equation}
can be solved, giving (for initial condition $\vt{B}(r,t',t')=\vt 1$)
 \begin{eqnarray}
   \vt{B}(r,t,t') =  \exp\left(\nabla\vb{v}\ (t-t')\right) = \vt 1 + \nabla \vb v (t-t') + \frac{1}{2!}\nabla\vt v \cdot \nabla \vt v (t-t')^2 + \dots
 \end{eqnarray}

 Therefore, taking up to first order terms in $(t-t')$, we can write 
  \begin{eqnarray}
   \int^{t_{0}}_{t_0-\Delta t_{cut}} M(t-t')\vt{\gamma}_{[0]}(t,t')dt'
    &\approx&\int^{t_{0}}_{t_0-\Delta t_{cut}} M(t-t') \nabla\vb{v}\ (t-t')dt',
   \nonumber\\
 \end{eqnarray}
 where $\Delta t_{cut}$ is the time interval during which we treat the velocity gradient tensor as constant. With the approximation of constant velocity gradient tensor, we can treat this tensor as constant during the time interval $(t_0-\Delta t_{cut}, t_0)$, thus
 \begin{eqnarray}
   \int^{t_{0}}_{t_0-\Delta t_{cut}} M(t-t')\vt{\gamma}_{[0]}(t,t')dt'
    &\approx&\nabla\vb{v}\int^{t_{0}}_{t_0-\Delta t_{cut}} M(t-t') (t-t')dt'.
    \nonumber\\
    \label{eq_constant_vgt1}
 \end{eqnarray}
 
By making use of equation \eqref{eq:memory_vs_relaxation}, we can integrate by parts to obtain
 \begin{eqnarray}
   \int^{t_{0}}_{t_0-\Delta t_{cut}} M(t-t')\ (t-t') dt' &=&
    -\Delta t_{cut}  G(\Delta t_{cut}) +
     \int ^{t_{0}}_{t_0-\Delta t_{cut}}G(t-t') dt',
     \nonumber\\
     \label{eq_constant_vgt2}
 \end{eqnarray}
 where the last integral yields
  \begin{eqnarray}
   \int^{t_{0}}_{t_0-\Delta t_{cut}} G(t-t') dt' &=&
     \mathbb{G} (\Delta t_{cut})^{-\beta + 1} E_{\alpha-\beta, 2-\beta}\left(-\frac{\mathbb{G}}{\mathbb{V}}
  (\Delta t_{cut})^{\alpha-\beta}\right).
  \nonumber\\
  \label{eq_constant_vgt3}
 \end{eqnarray}
%
For additional information on how the integration is carried out in the code, please refer to \ref{appendix_integration}.\\
%
%
We highlight that in this formulation we kept the general assumption on $\boldsymbol{\gamma}_{[0]}$, therefore we did not restrict ourselves to the case of small deformations. While the results are equivalent under SAOS, this approach allows us to eventually extend the scheme to include large amplitude oscillations or more complex flows in the future.

\section{Numerical simulations}

\begin{figure}[bt]
    \centering  
    \includegraphics[width=0.8\linewidth]{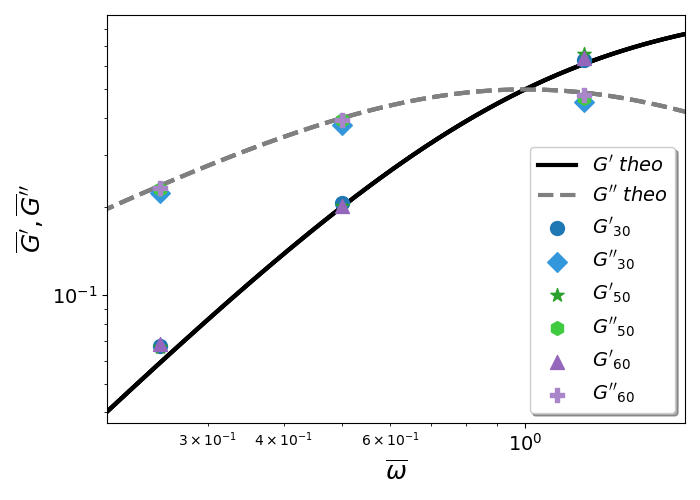}
    \caption{Effect of the spatial resolution, storage and loss moduli vs frequency. Results improve when a larger number of particle is chosen, and converge at around $50\time50$ particles. Black solid line and grey dashed line for analytical $G'$ and $G''$ respectively. The subscripts indicate the number of particles. $30\times30 \to$ \textcolor{blue}{$\bullet$}:  $G'$, \textcolor{blue}{$\blacklozenge$}:  $G''$. $50\times50 \to$  \textcolor{green}{$\star$}:  $G'$, \textcolor{green}{$\varhexagonblack$}: $G''$. $60\times60 \to$  \textcolor{violet}{$\blacktriangle$}:  $G'$, \textcolor{violet}{$+$}: $G''$.  }
    \label{fig:spatialResolution}
\end{figure} 

\subsection{Numerical results}
\begin{figure}[tb]
    \centering
    \includegraphics[width=0.8\linewidth]{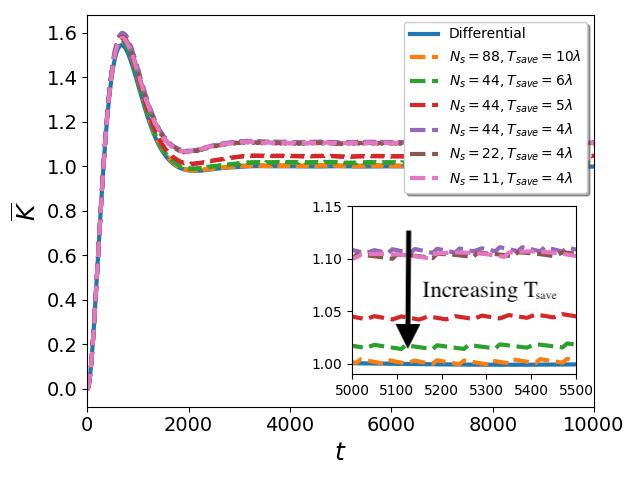}
    \caption{Kinetic energy $\overline{K}$ vs time $t$ for a Kolmogovov flows. Comparison for differential (solid line) and integral (dashed lines) simulations at various time-resolution parameters $T_{save}$, $N_s$. In the inset, zoom of steady flow. Results converge to the expected value when a longer $T_{save}$ is chosen.}
    \label{fig:test-kolmogorov}
\end{figure}
\label{sec:results}
\subsubsection{Small Amplitude Oscillatory Shear (SAOS)}
In this section we verify that the integral approach to the evolution produces correct results. In order to do so,  we focus on simulating a viscoelastic fluid under Small Amplitude Oscillatory Shear (SAOS), i.e. a fluid confined between two parallel plates posed at a distance $H$, with the bottom plate fixed and the top plate oscillating at a frequency $\omega$. The corresponding transient shear-strain/rate read, respectively, 
\begin{align}
    &\gamma(t)=\gamma_0 \sin(\omega t) \\
    &\dot\gamma(t)=\gamma_0\omega \cos(\omega t) 
\end{align}
being $\gamma_0=\Delta/H$ the ratio between the maximum displacement of the upper wall $\Delta$ and the height.
Under SAOS, it is common to focus on the complex relaxation modulus $G^*(\omega) = G'(\omega) + iG''(\omega)$, with $i=\sqrt{-1}$, where $G'(\omega)$ is the elastic modulus and $G''(\omega)$ the viscous modulus. For a pure Newtonian fluid, at low frequencies the dominant contribution is $G^*_N(\omega)=iG''(\omega)\approx i\eta_s\omega$; however, at higher frequencies the inertial effects become non-negligible and produce a non-zero effective elastic modulus \cite{boehme1990influence, villone2019numerical}, obtaining
\begin{equation}
    G^*_{N}(\omega) = i \eta_s \omega \frac{\sqrt{\frac{\rho\omega H^2}{\eta_s} i}}{\sinh(\sqrt{\frac{\rho\omega H^2}{\eta_s}i})}
    \label{eq:G_newtonian}
\end{equation}
for which the real part scales quadratically with the frequency. Therefore, for viscoelastic fluids, we must correct the measured $G', G''$ by considering  this additional apparent elastic term  in the contribution derived by viscoelasticity.

For a Maxwell fluid with 
polymeric viscosity $\eta_p$ and elastic relaxation time $\lambda$, the non-inertial viscoelastic component is \cite{bird1987dynamics}
\begin{align}
    &G'(\omega) = G_0\frac{\omega^2\lambda^2}{\omega^2\lambda^2+1}\\
    &G''(\omega) =G_0\frac{\omega\lambda}{\omega^2\lambda^2+1}
\end{align}
with $G_0={\eta_p}/{\lambda}$, while for a Fractional Maxwell (FMM) model we have \cite{rathinaraj2021incorporating}
\begin{align}
    G^*(\omega)=\frac{\mathbb G(i\omega)^\beta\cdot\mathbb V(i\omega)^\alpha}{\mathbb G(i\omega)^\beta+\mathbb V(i\omega)^\alpha}
\end{align}
which results in
\begin{align}
    G'(\omega) &= \frac{ \qty(\mathbb G \omega^\beta)^2\mathbb V \omega ^\alpha \cos(\pi\alpha/2) + \qty(\mathbb V \omega^\alpha)^2\mathbb G \omega ^\beta \cos(\pi\beta/2) } {\qty(\mathbb G \omega^\beta)^2 + \qty(\mathbb V \omega^\alpha)^2 + 2\mathbb G \omega^\beta \mathbb V \omega^\alpha \cos(\pi(\alpha-\beta)/2)}\\
    G''(\omega) &= \frac{ \qty(\mathbb G \omega^\beta)^2\mathbb V \omega ^\alpha \sin(\pi\alpha/2) + \qty(\mathbb V \omega^\alpha)^2\mathbb G \omega ^\beta \sin(\pi\beta/2) } {\qty(\mathbb G \omega^\beta)^2 + \qty(\mathbb V \omega^\alpha)^2 + 2\mathbb G \omega^\beta \mathbb V \omega^\alpha \cos(\pi(\alpha-\beta)/2)}    
\end{align}
A characteristic relaxation time for the FMM can be defined as \cite{jaishankar2014fractional} 
\begin{equation}
\lambda_{F}=\qty(\frac{\mathbb V}{\mathbb G})^\frac{1}{\alpha-\beta}
\end{equation}
and a characteristic modulus is identified by $G_0=\mathbb V \lambda_{F}^{-\alpha}$.

To extract the storage and loss moduli, the following procedure is adopted. 
 The off-diagonal component of the stress tensor  $\vb \tau\equiv\vt \tau_{xy}$ is related to $G'(\omega)$ and $G''(\omega)$ via 
 \begin{align}
     &\tau(t)=\frac{G''(\omega)}{\omega}\dot\gamma(t)+G'(\omega)\gamma(t)=\frac{\dot\gamma_0}{\omega}G(\omega)\sin(\omega t + \delta(\omega))\\
     &G'(\omega)=G(\omega)\cos(\delta(\omega))\\
     &G''(\omega)=G(\omega)\sin(\delta(\omega))
 \end{align}
Assuming that $\tau$ is constant in the domain (i.e. linear SAOS regime), plotting a Lissajous curve (an ellipse for a viscoelastic material under SAOS) of $\dot\gamma(t)$ versus $\vb \tau(t)$ allows us to extract $G$ and $\delta$ as the best fitting parameters. The full procedure is described in details in \cite{vazquezquesada2012sph}.
\begin{figure}[tb]
    \centering
    \includegraphics[width=1.0\linewidth]{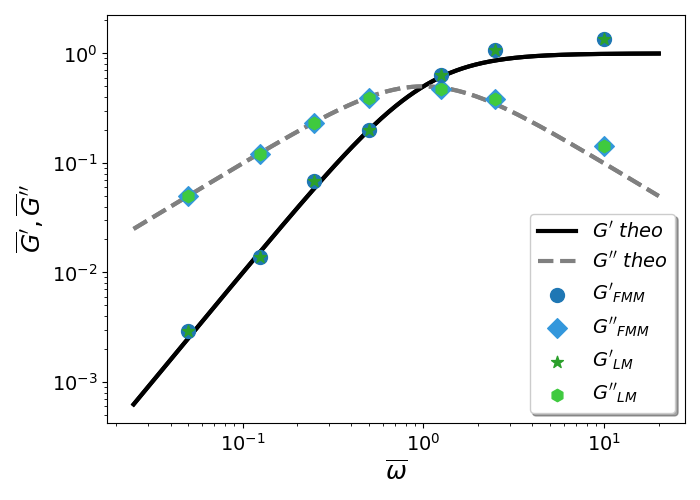}
    \caption{SAOS analysis (storage and loss moduli vs frequency) for FMM with $\alpha=1, \beta=0, \mathbb V=450, \mathbb G=900$.  Results are equivalent to a Linear Maxwell model with $\lambda=0.5, \eta_p=900$. Black solid line and grey dashed line for analytical $G'$ and $G''$ respectively. Simulated Fractional Maxwell Model (FMM): \textcolor{blue}{$\bullet$}:  $G'$. \textcolor{blue}{$\blacklozenge$}:  $G''$. Simulated Linear Maxwell (LM): \textcolor{green}{$\star$}:  $G'$. \textcolor{green}{$\varhexagonblack$}: $G''$ }
    \label{fig:comparisonMaxwell}
\end{figure}

Given the symmetry of the problem, we can study the system performing two-dimensional simulations. The flow domain is a square $[0 : L]\times[0 : L]$, $L=7.5$, with no-slip boundary conditions at the channel walls for $y=\{0,L\}$ and periodic boundary conditions on the flow direction at $x=\{0,L\}$. The upper wall oscillates as $V_t=V_{max}\sin(\omega t)$, with $V_{max}(\omega)=\omega\Delta/L$, with $\Delta=0.05L$ being the maximum displacement, small enough to keep the linear regime of SAOS. The number of fluid particles in the system is $N_p=N\times N$ (since the domain is a square, we use the same number of particles in the two dimensions). Spatial resolution has been verified by running simulations at various $N$, fixing the cut-off radius to $4L/N$. In figure \ref{fig:spatialResolution} we have verified that the simulation converges to the expected result for $N\ge50$, therefore we have chosen to perform the following analysis with $N = 60$ and a cut-off radius of $r_c=0.5$. Overlined quantities represent normalised quantities: $\overline{\omega} = \omega \min(\lambda_k)$ or $\omega\lambda_F$ for multi-Mode Maxwell and Fractional Maxwell models respectively; $\overline{G} = G/G_{0}$, with $G_0 = \min(\eta_k/\lambda_k)$ for multi-Mode Maxwell model and $G_0=\mathbb V \lambda^{-\alpha}$ for Fractional Maxwell model.
\begin{figure}[tb]
    \centering
    \includegraphics[width=1.0\linewidth]{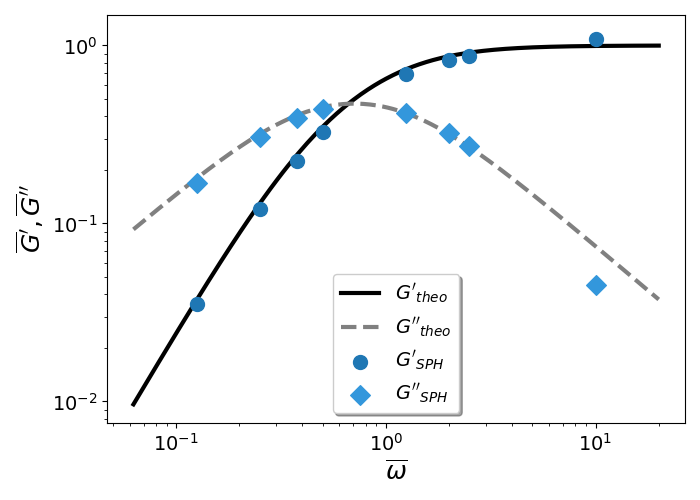}
    \caption{SAOS analysis (storage and loss moduli vs frequency) for multi-mode Maxwell with $\lambda_1=0.5$, $\lambda_2=1.0$ and $\eta_{1}=\eta_{2}=900$. Black solid line and grey dashed line for analytical $G'$ and $G''$ respectively. Simulation data: \textcolor{blue}{$\bullet$}:  $G'$. \textcolor{blue}{$\blacklozenge$}:  $G''$. }
    \label{fig:multimode}
\end{figure}
For what concerns the other parameters, in the Maxwell analysis the solvent and polymeric viscosity are equal to $\eta_s=\eta_p=900$, $\rho=1$, the speed of sound is $c_s=10\gg V_{max}$ so that the Mach number $V_{max}/c_{s}$ is always smaller than $1$, thus avoiding liquid compressibility effect. The elastic time $\lambda=0.5$ is chosen to satisfy the series of disequalities $t_\eta = \frac{L^2\rho}{\eta_s+\eta_p} < \lambda < t_{\omega}=\frac{2\pi}{\omega}$ \cite{vazquezquesada2012sph}. Given a time-step $\Delta t\approx 10^{-4}\lambda$, we choose $T_{save} = 10\lambda$,  and the integral \eqref{eq:memory_integral}  is performed over $N_s=88$ flow configurations: the accuracy of this approximation is shown in figure \ref{fig:test-kolmogorov}, where results of a classical Kolmogorov test are shown, i.e. periodic body force in the $x$ direction, denoted as $\vb F=(F_0\sin(k y),0)$ whith $k = 2\pi/L$, is applied to a fluid within a simulation box featuring periodic boundary conditions in all directions \cite{vazquezquesada2009smoothed}. This system is analyzed for various time-resolution parameters. The results are normalized using $\overline{K}=K/K_{\text{steady}}$, where $K$ represents the total kinetic energy of the system, and $K_{\text{steady}}$ is the total kinetic energy of the system at the steady state. In the figure the results from several simulations with different values of $N_s$ and $T_{save}$ are compared with the result obtained from a simulation, with the same parameters, of an Oldroyd-B model with a differential code \cite{vazquezquesada2009smoothed}. The graph shows that the results of the integral model converges toward the differential result when both $N_s$ and $T_{save}$ values are increased.

In figure \ref{fig:comparisonMaxwell} we show  SAOS simulations done over two orders of magnitude of frequency, comparing results between the Fractional Maxwell Model and the Linear Maxwell model. Fractional simulations are carried out in the limit $\alpha=1, \beta=0$ to recover the same behaviour of the Linear Maxwell. The chosen memory functions for these simulations are, respectively, eq. \eqref{eq:memory_fractional} with parameters  $\alpha=1, \beta=0, \mathbb V=450, \mathbb G=900$ and eq. \eqref{eq:memory_singleMaxwell} with parameters $\lambda=0.5, \eta_p=900$. Numerical results show a perfect agreement with theoretical values, which is only partially lost at higher frequencies due to the fact that, for very large $\omega$, non-linear effects start to emerge.


We then explore more complex constitutive equations. First, we analyse the behaviour of a multi-mode Maxwell mode, using the memory function of equation \eqref{eq:memory_multiMaxwell}. The chosen parameters for this two-modes simulation are $\lambda_1=0.5, \lambda_2=1, \eta_{1}=\eta_{2}=900$. Results for this study are shown in figure \ref{fig:multimode}, where a very good agreement with theory is observed. 

\begin{figure}[tb]
    \centering
    \includegraphics[width=1.0\linewidth]{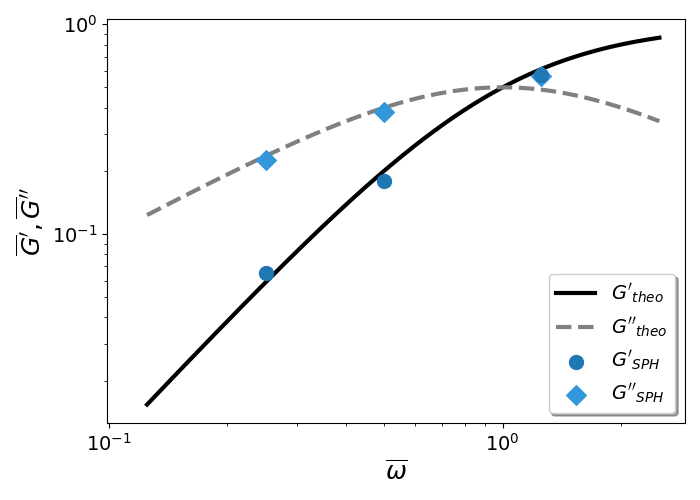}
    \caption{SAOS analysis (storage and loss moduli vs frequency) for Fractional Maxwell with $\alpha=1, \beta=0.1, \mathbb V=450, \mathbb G=900$ Black solid line and grey dashed line for analytical $G'$ and $G''$ respectively. Simulation data: \textcolor{blue}{$\bullet$}:  $G'$. \textcolor{blue}{$\blacklozenge$}:  $G''$. }
    \label{fig:realFractional}
\end{figure}

For the fractional Maxwell model, the results of SAOS simulations done at $\alpha=1, \beta=0.1, \mathbb V=450, \mathbb G=900$ are shown in figure \ref{fig:realFractional}. A good agreement between the theoretical prediction and the experimental data is present in the range of frequencies explored here.

\section{Conclusions}
\label{sec:conclusions}
Integral constitutive equations were introduced decades ago and are particularly effective in describing some complex fluids, such as polymer melts. However, their numerical implementation is complicated by the requirement of tracking the particle history along the whole domain, which can be extremely hard to achieve. Using the SPH framework, we exploited its property of naturally tracking the particle position to discretize integral schemes in an efficient way. We focused on the analysis of polymer melts under SAOS. This kind of material present a multi-scale power law behaviour in the relaxation modulus, which can be captured by either a significantly large amount of modes in a multi-mode Maxwell model, or by the introduction of fractional constitutive equations. We have shown how our scheme can easily implement both approaches by taking advantage of the structure of the integral model, and verified that numerical simulations are in very good agreement with theoretical results. 

While we simulated flows with linear deformations, we built our model without the assumption of linearity, therefore our numerical scheme can be easily expanded to include non-linear behaviour as well in a future development. In particular, the model we are employing can be easily extended to more complex flows, such as large amplitude oscillatory shear (LAOS) or other flows with mixed shear/extensional types. In order to do so, it is important to allow for non linearity in the constitutive equation.
Equation \eqref{eq:memory_integral} is often called a quasi-linear constitutive equation, because the stress depends linearly on the strain, but the strain is non linear in the displacement. A more complete expansion is the K-BKZ model, which under some assumptions in a shear deformation reads $ \tau = \int M(t-t')h(\vt \gamma_{[0]}(t,t'))\vt\gamma_{[0]}(t,t') dt'$, with $h(\vt\gamma_{[0]}(t,t'))$ damping function to be determined independently \cite{bird1987dynamics,owens2002computational, jaishankar2014fractional}. This constitutive equation can be built effortlessly in the scheme we propose here, and will be the subject of future investigations on the behaviour of multi-scale materials.

\section{Acknowledgements}
This work has been partially funded by the Basque Government through
the ELKARTEK 2022 programme (KAIROS project: grant KK-2022/00052). Financial support through the BERC 2022-2025 program and by the Spanish State Research Agency through BCAM Severo Ochoa Excellence Accreditation CEX2021-001142-S/MICIN/AEI/10.13039/501100011033 and through the projects PID2020-117080RB-C55 (‘Microscopic foundations of soft-matter experiments: computational nano-hydrodynamics’, with acronym ‘Compu-Nano-Hydro) and PID2020-117080RB-C54 (‘‘Coarse-Graining theory
and experimental techniques for multiscale biological systems’’) funded both by AEI – MICIN are also gratefully acknowledged. UNED funding for open access publishing is also acknowledged.

\appendix

\section{Details of the integration for the calculation of the stress tensor}
\label{appendix_integration}

\begin{figure}
    \centering
    \includegraphics[width=1.0\linewidth]{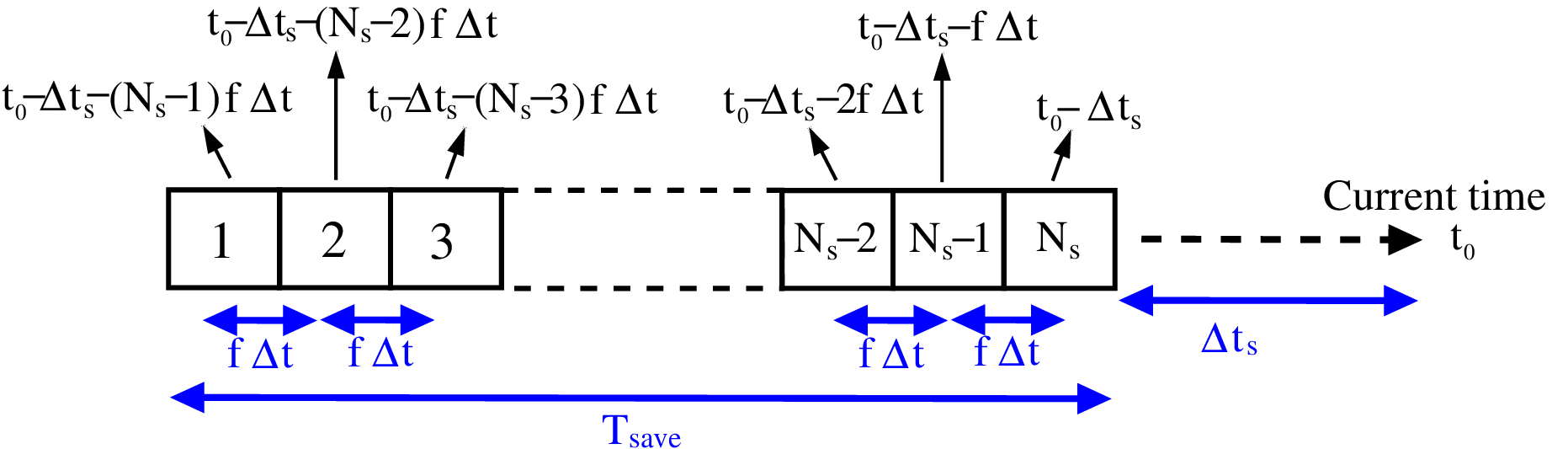}
    \caption{\textcolor{black}{Explanatory diagram of the array used in the program. The displacements
for $N_s$ different time intervals are stored in an array for each particle. The time interval
between two consecutive positions is $f\Delta t$. The times corresponding to the positions are
indicated in the diagram. It is important to note that at each time step, $\Delta t_s$ is changing, so the corresponding time for each position in the array is also changing.}
    }
    \label{Diagram_array}
\end{figure}

\textcolor{black}{
The diagram in figure \ref{Diagram_array} schematically illustrates the algorithm used to calculate the stress tensor of particle $i$. Eq. (\ref{stress_calculation}) is employed for this purpose, involving an integration of the memory function multiplied by the relative finite strain tensor. The memory function $M(t-t')$ depends on the selected time interval, as indicated in Equation (\ref{eq:memory_fractional}), while $\boldsymbol{\gamma}_{[0],i}(t,t')$ can be calculated from the particle positions at a previous time $t'$, as described in Equation (\ref{relative_finite_stress_tensor}).}

\textcolor{black}{
Particle positions are recorded at intervals of $f\Delta t$, where $\Delta t$ represents the time step. It is not necessary to store positions for every time step since we expect the curve $M(t-t')\boldsymbol{\gamma}_{[0],i}(t,t')$ to be smooth. To achieve this, an array with $N_s = T_{save}/(f\Delta t)$ positions is assigned to each fluid particle to store their past positions, as shown in figure \ref{Diagram_array}. In this arrangement, the position $l$ in the array corresponds to the particle's location at time $t_0-\Delta t_s - (N_s - l)f\Delta t$, where $\Delta t_s$ is the time interval between the current time $t_0$ and the most recent time when the particle positions were stored. It's worth noting that with this array structure, the most recent past position of particle $i$ is stored at location $N_s$, and the oldest past position is stored at position $1$. The time interval $\Delta t_s$ is variable, increasing $\Delta t$ at each time step. Consequently, the time corresponding to every element of the array also changes.
Initially, $\Delta t_s$ is zero. After every $f$ time steps, the array needs to be updated, with all stored positions at locations $l>1$ being moved from $i$ to $l-1$, and the position at location $N_s$ initialized with the current position of the fluid particle.}

\textcolor{black}{
As explained in section \ref{section_integral_compute}, when using periodic boundary conditions, it is simpler to store the displacements of the particle with respect to previous times rather than the positions. This is because even though displacements need to be updated at each time step with Equation (\ref{update_displacement}), it is straightforward to add or subtract the box's length when a particle crosses the simulation box's boundaries.}

\textcolor{black}{
With respect to the calculation of the stress tensor for particle $i$, according to Eqs. (\ref{stress_calculation}) and (\ref{eq_constant_vgt1}), the integration to calculate the stress of particle $i$, if we consider that its velocity gradient tensor is changing slowly, is as follows:
\begin{eqnarray}
\boldsymbol{\tau}_i = 
\int_{t_0-T_{save}}^{t_0-\Delta t_{cut}} M(t-t')
\boldsymbol{\gamma}_{[0],i}(t,t') \dd t'
+ \left(\nabla \boldsymbol{v}\right)_i
\int_{t_0-\Delta t_{cut}}^{t_0} (t-t')M(t-t') \dd t' \
\
\
\end{eqnarray}
Given that the relative strain tensor $\boldsymbol{\gamma}_{[0],i}(t,t')$ can be calculated from the states of the particle that have been stored at various points in the past, the first integral can be numerically evaluated. In our case, we have employed a composite Simpson's $1/3$ rule. The reason for not using this numerical scheme for the entire history of the particle is that the integrand can be divergent for the fractional case when $t'\rightarrow t_0$. Hence, it is more convenient to compute the integral analytically. Furthermore, even in cases where this quantity is not divergent, the time interval $\Delta t_s$ between the last saved point and the current time is generally different from the frequency $f\Delta t$ used to store the rest of the history. A Simpson's rule-type algorithm would yield inaccurate results when $\Delta t_s$ is very small or when it is close to $f\Delta t$ since these cases involve interpolations with points that are very close together. To circumvent these issues, the second integral is calculated analytically using the expressions (\ref{eq_constant_vgt2}) and (\ref{eq_constant_vgt3}) with $\Delta t_{cut} = \Delta t_s + f\Delta t$.
}
\clearpage

\bibliographystyle{unsrt} 
\bibliography{SPH_2023nonNewtonian} 
\end{document}